\documentstyle[aps,epsf]{revtex}

\global\firstfigfalse  % kludge to bypass foolishness of revtex to place
\global\firsttabfalse  % figures and tables at end of paper

\def \chizero  {\tilde{\chi}_1^0}
\def \chitwo   {\tilde{\chi}_2^0}
\def \chione   {\tilde{\chi}_1^\pm}
\def \ET       {E_{T}}
\def \MET      {\mbox{$\raisebox{.3ex}{$\not$}\ET$}}
\def \ppb      {\mathrm{p\bar p}}
\def \lnu      {\ell\nu}
\def \pgev     {\mathrm{GeV}/{\it{c}}}
\def \bbb      {\mathrm{b\bar b}}
\def \ccb      {\mathrm{c\bar c}}
\def \ttb      {\mathrm{t\bar t}}
\def \ee       {\mathrm{e^+e^-}}

\def \PT       {P_{T}}
\def \mhalf    {m_{1/2}}
\def \pbinv    {\mathrm{pb}^{-1}}
\def \Gcsq     {\mathrm{GeV}/{\it{c}}^2}
\def \mzero    {m_{0}}
\def \tanb     {\tan\beta}

\begin{document}

\baselineskip 14pt

%\begin{flushright}
%CDF/ANAL/EXOTIC/CDFR/4828 \\
%\today\\
%\end{flushright}

\title{Study of a Like-Sign Dilepton Search for Chargino-Neutralino 
Production at CDF}
\author{Jane~Nachtman, David~Saltzberg, Matthew~Worcester
\footnote{nachtman@physics.ucla.edu, saltzbrg@physics.ucla.edu, 
mworcest@physics.ucla.edu}}
\address{for the CDF Collaboration \\
Department of Physics and Astronomy, University of California, Los Angeles}

\maketitle     

\begin{abstract}

We propose a like-sign dilepton search for chargino-neutralino 
production in $\ppb$ collisions at $\sqrt{s} = 1.8$ TeV, which 
complements the previously published trilepton search
by the CDF detector using Fermilab Run I data.
Monte Carlo predictions for the signal and background efficiencies
indicate a significant increase in sensitivity to chargino-neutralino 
production compared to the traditional trilepton analysis alone.

\end{abstract}

\section{Introduction}

Previous searches for chargino-neutralino production at the Tevatron have 
focused primarily on signatures with three charged leptons (trileptons) plus 
missing  transverse energy ($\MET$)~\cite{trilep}.  In the Minimal 
Supersymmetric (SUSY) Standard Model, chargino-neutralino production occurs 
in proton-antiproton ($\ppb$) collisions via a virtual W (s channel) or a 
virtual squark (t channel).  In a representative minimal Supergravity 
(SUGRA) model (parameters: 
$\mu < 0, \tanb = 2, \mathrm{A_{0}} = 0, \mzero = 200~\Gcsq$, 
$\mhalf = 90-140~\Gcsq$
), we expect three-body chargino and neutralino decays through virtual 
bosons and sleptons in a chargino mass region of $80-130~\Gcsq$.  Conserving 
R-parity, these decays produce a distinct signature: trileptons plus
$\MET$ from a neutrino and the lightest supersymmetric
particle.  We demonstrate that the sensitivity to this signature can be
significantly increased by searching for events with two like-sign leptons.  
The Like-Sign Dilepton (LSD) search provides a strong rejection 
of Standard Model background through the like-sign requirement, and enhances 
the acceptance of the signal by requiring only two of the three leptons
produced in the chargino-neutralino decay.

\section{Like-Sign Dilepton Analysis}

Signal and most background processes were generated using ISAJET 7.20 and 
the CDF detector Monte Carlo simulation.  For the signal estimation, we used 
representative SUGRA parameters of 
 $\mu < 0, \tanb = 2, \mathrm{A_{0}} = 0, \mzero = 200~\Gcsq$, and
 $\mhalf = 90-140~\Gcsq$.  The relevant mass relations are  
 $\mathrm{M_{\chione}} \sim \mathrm{M_{\chitwo}} \sim 2\mathrm{M_{\chizero}}$
, with
 $\mathrm{M_{\chione}}$ between $80-130~\Gcsq$.  The sleptons and sneutrinos
have masses between $200-220~\Gcsq$, so we generate only three-body
chargino and neutralino decays.

The LSD analysis begins with the selection of a pair of leptons ($ee, \mu\mu, 
e\mu$) with the same charge.  We then impose kinematic requirements 
on the selected events in order to
remove Standard Model and other non-SUSY backgrounds.  Our primary requirements
are minimum transverse momentum ($\PT > 11~\pgev$) for both leptons, 
and isolation, in which we remove events where at least one lepton
has excess transverse energy greater than 2 GeV in a cone of 0.4 radians 
around the lepton.  Monte Carlo simulations indicate that isolation 
removes heavy flavor ($\bbb,\ccb$) backgrounds most effectively.  
As the like-sign cut requires us to select both leptons from a b or c decay 
in such an event, and as semi-leptonic b and c decays produce 
leptons associated with jets, neither of the selected leptons will 
be isolated.  Isolation also reduces $\ttb$ because at least one lepton from
the like-sign pair will be selected from a b decay in such an event.  
The isolation cut, when applied to both like-sign leptons, 
reduces $\bbb$ and $\ccb$ to a negligible level.

We remove diboson events 
through a Z-mass rejection, in which the combined mass of a third 
opposite-sign, same-flavor lepton selected by the analysis and either 
of the LSDs is between $80-100~\Gcsq$, reducing WZ and ZZ backgrounds.  
We impose no requirement on $\MET$.  This leaves WZ production as the 
dominant source of Standard Model background, as shown in Table 1.

An important source of non-SUSY background estimated from data is events with 
one true lepton, such as W $\rightarrow \lnu$ + jets, and a ``fake'' 
lepton, {\it{i.e.}} an isolated track misidentified as a lepton.
This fake lepton, in combination with the true lepton from 
the W decay, can be selected as a signal event in this analysis.  In 
order to estimate this background, we first look at Z $\rightarrow \ee$  
+ jets, which we assume provides a model for W + jets events.  
Removing the true leptons, we then measure the rate of underlying isolated 
tracks in the event.  Next we search minimum bias data, 
in which we assume there are no true leptons, to find the probability of an
isolated track to be misidentified as a lepton.  The probability
of misidentifying an isolated track as a lepton is 1.5$\%$ per track.
We multiply this probability by the isolated track rate from the Z 
$\rightarrow \ee$ events, by the number of W + jets events 
expected~\cite{run1w}, and by a factor of 0.5 for the like-sign requirement.  
This ``fake'' rate drops rapidly with an increasing minimum $\PT$ 
requirement.  Optimization of the number of expected background events 
as a function of the $\PT$ requirement yields 0.3 events expected from 
W + jets in 100~$\pbinv$ of data.

\section{Results}

Applying the analysis requirements and normalizing the luminosity to
100~$\pbinv$, the expected background is a total of 0.56
events, as shown in Table 1.  Drell-Yan and W + jets are the most 
significant non-SUSY backgrounds; WZ production is the largest Standard
Model background. There is little background overlap of the 
trilepton and LSD analyses in the selected events based on Monte Carlo
studies.  Therefore, the backgrounds are treated as independent.  For 
the trilepton analysis, the expected background for the Run I luminosity
of 107~$\pbinv$ is 1.2 events~\cite{trilep}.  The total expected background 
for the combined LSD and trilepton analyses is 1.8 events.

\begin{table}

\caption{Background estimates for the number of events
expected in 100~$\pbinv$ of data based on Monte Carlo (except for the 
W + jets data estimation).  The errors are one-sigma statistical errors.}

\begin{tabular}{|c|c|l|}

Process   &   Luminosity($\pbinv$)    & $\mathrm{N_{events}}$ expected \\ 
\tableline
\tableline
WZ 	&	16,684		&  $0.11\pm 0.02$ \\ 
\tableline
ZZ      &	13,992		&  $0.01\pm 0.01$ \\ 
\tableline
WW      &        6,870		&  $0^{+0.02}_{-0}$ \\ 
\tableline
$\mathrm{t\bar{t}}$ & 5,558     &  $0^{+0.02}_{-0}$ \\ 
\tableline
Drell-Yan($\gamma^{*}/Z$)  &        1,728       &  $0.11^{+0.10}_{-0.06}$ \\ 
\tableline
$\mathrm{\bbb,\ccb}$   & 3,122     &   $0.03^{+0.04}_{-0.02}$ \\ 
\tableline
W + jets  &  (from data)         &    0.30  \\ 
\tableline 
\tableline
Total     &                     &    0.56

\end{tabular}

\end{table}

\begin{figure}[ht]
	
\centerline{\epsfxsize 5.0 truein \epsfbox{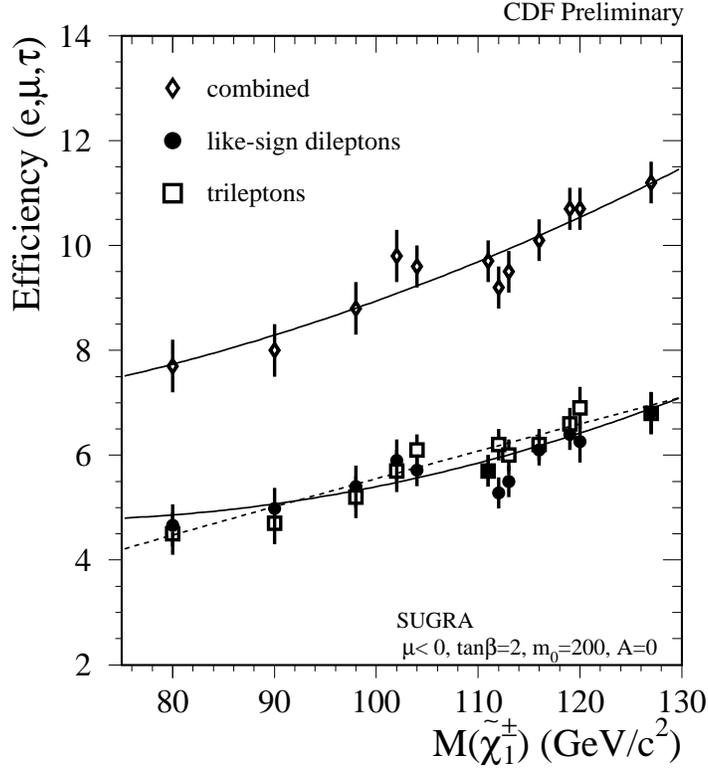}}   
\vskip -.2 cm
\caption[]{
\label{eff}
\small The efficiency (as the percentage of events selected) for the 
trilepton, LSD, and combined analyses as a function of chargino mass.  
SUGRA parameters of 
 $\mu < 0, \tanb = 2, \mathrm{A_{0}} = 0, \mzero = 200~\Gcsq$, and
 $\mhalf = 90-140~\Gcsq$ were used to measure the efficiency.}

\end{figure}

Figure~\ref{eff} shows the efficiency versus
chargino mass for the trilepton analysis, the LSD analysis, and the combined
analyses, taking into account the signal overlap between the trilepton and LSD 
analyses.  These efficiencies are calculated for all three analyses as
number of selected events divided by total number of chargino-neutralino 
events where both sparticles decay leptonically, where a lepton
can be $e, \mu,$ or $\tau$.  All $\tau$ decays are included in this 
calculation, even though the analyses are only sensitive to the 
leptonic decays.

\begin{figure}[ht]

\centerline{\epsfxsize 5.0 truein \epsfbox{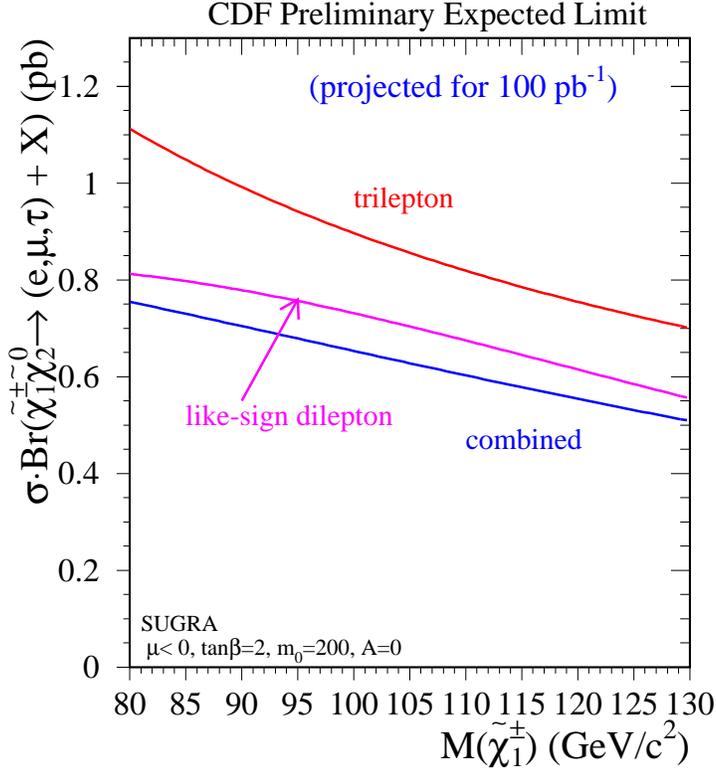}}   
\vskip -.2 cm
\caption[]{
\label{explim}
\small Average expected limit on $\sigma~\cdot~\cal{B}$ as a function 
of M$_{\chione}$ for the LSD analysis, trilepton analysis, and the combination
of both analyses.  The parametrized efficiencies shown in 
Figure~\ref{eff} are used along with the expected background from
Monte Carlo in this calculation.}

\end{figure}

Figure~\ref{explim} shows the average expected limit normalized to 
100~$\pbinv$ for the trilepton, LSD, and combined analyses.  These limits 
were calculated from the efficiencies in Figure~\ref{eff} and from the 
expected number of background events based on Monte Carlo.

\section{Conclusion}

This study indicates that a fully realized Like-Sign
Dilepton analysis will increase the sensitivity of searches 
for chargino-neutralino production with the CDF detector using existing data 
of $\ppb$ collisions at $\sqrt{s} = 1.8$ TeV.  
It has been shown that the sensitivity of the previously published  
trilepton analysis can be improved by combining it with this new
LSD signature search.  Significantly, the LSD search has fewer requirements 
than the trilepton analysis, {\it{e.g.}} the trilepton analysis requires 
$\MET > 15~\Gcsq$ whereas the LSD analysis has no $\MET$ requirement, 
making the Like-Sign Dilepton channel sensitive to a greater number of 
signatures.

This research was supported by a grant (number DE-FG03-91ER40662) from the 
U.S. Department of Energy.

\end{document}